\documentclass[%
 reprint,
superscriptaddress,
 amsmath,amssymb,
 aps,
]{revtex4-2}

\usepackage[utf8]{inputenc}
\usepackage{amsmath}
\usepackage{amsfonts}
\usepackage{braket}
\usepackage{comment}
\usepackage[normalem]{ulem}
\usepackage{physics}
\usepackage{xcolor}
\usepackage{ulem}
\usepackage{graphicx}% Include figure files
\usepackage{dcolumn}% Align table columns on decimal point
\usepackage{bm}% bold math
\usepackage{hyperref}% add hypertext capabilities

\begin{document}

\preprint{APS/123-QED}

\title{A Collective Propagation Law of Optical Vortex Constellations and Longitudinal Sensing}

\author{Niladri Modak}
\email{niladri.modak@tuni.fi}
\affiliation{Tampere University, Photonics Laboratory, Physics Unit, Tampere, FI-33720, Finland}

\author{Rafael F. Barros}
\affiliation{Instituto de Física, Universidade de São Paulo, 05315-970 São Paulo, SP, Brazil}

\author{Marco Ornigotti}
\affiliation{Tampere University, Photonics Laboratory, Physics Unit, Tampere, FI-33720, Finland}

\author{Robert Fickler}
\affiliation{Tampere University, Photonics Laboratory, Physics Unit, Tampere, FI-33720, Finland}

\date{\today}

\begin{abstract}
Optical vortices are ubiquitous phenomena naturally appearing in wave physics, yet higher-order charges inherently split into constellations of lowest-order singularities at the smallest deviation from an ideal situation. 
While these constellations are common phenomena in real-world scenarios, characterizing their longitudinal evolution typically relies on exhaustive full-field descriptions or the ambiguous, sequential tracking of individual singularities.
Here, we reveal and experimentally demonstrate a simple deterministic law describing the paraxial longitudinal propagation of an arbitrary constellation of optical vortices in standard Gaussian backgrounds. 
By mapping the constituting singularity coordinates to their elementary symmetric polynomials (ESPs), we capture the holistic evolution of the constellation during propagation, completely bypassing the practical need to sequentially track indistinguishable vortices.
We further show that such complex ESPs provide a useful metrological tool for the estimation of longitudinal displacements.
Our results reveal a previously unrecognized compact description of collective vortex dynamics, introducing a new route to longitudinal sensing through singularimetry.

\end{abstract}

\maketitle

%\tableofcontents

%%%%%%%%%%%%%%%%%%%%%%%%%%%%%%%%%%%%%%%%%%%%%%%%%%%%%%
\noindent \emph{Introduction -} An optical vortex is a point in the complex electric field of light at which the field becomes undefined. 
A widely studied example is a phase vortex arising from a transverse helical phase structure of the form $\exp(i\ell\varphi)$, where $\varphi$ denotes the azimuthal angle and $\ell$ is the topological charge \cite{allen1992orbital,wang2016advances,shen2019optical,gbur2016singular,dennis2009singular,quinteiro2022interplay}. Such a twisted phase front gives rise to an optical vortex along the beam axis and is associated with an orbital angular momentum of $\ell\hbar$ per photon.
Optical vortex beams with higher topological charges have attracted sustained interest for addressing fundamental questions in optics, including spin--orbit interaction of light \cite{bliokh2015spin}, the rotational Doppler effect \cite{lavery2013detection,emile2023rotational}, and high-dimensional quantum entanglement \cite{mair2001entanglement,malik2016multi}.
Alongside, they have enabled a wide range of applications in microscopy and imaging \cite{hell1994breaking}, optomechanical manipulation \cite{he1995direct}, classical and quantum communications \cite{willner2021orbital}, and photonic quantum information processing \cite{brandt2020high}.

Ideally, a vortex of topological charge $\ell$ corresponds to a single phase singularity of order $\ell$. 
In practice, however, higher-order optical vortices are intrinsically unstable, even under small perturbations and experimental imperfections. 
As a result, any realistic implementation of an $\ell$-order vortex generically breaks up into a constellation of $n$-spatially separated unit-charge singularities (with $n=|\ell|$) concentrated around the beam center \cite{freund1999critical,ricci2012instability,neo2014correcting,dennis2012topological,barros2024observation}. 
Consequently, the experimentally relevant objects in singular optics are not isolated higher-order vortices, but a collection of unit-charge singularities forming an optical vortex constellation. 
Understanding and characterizing the behaviour, specifically, the longitudinal propagation of vortex constellations is a prerequisite for the effective use of such higher-order vortex fields as the longitudinal propagation is the most ubiquitous form of optical evolution in all experimental contexts.  
Existing approaches typically rely on exhaustive full-field descriptions or the sequential tracking of individual singularities, i.e., keeping a record about the trajectory of each individual vortex, where the mathematical complexity scales poorly with increasing number of singularities rendering them impractical \cite{indebetouw1993optical,christou1996vortex,rozas1997propagation,kivshar1998dynamics,molina2001propagation,flossmann2005propagation,baumann2009propagation,liu2025observation}.
Consequently, there is a critical need for a holistic mathematical framework that bypasses tracking of individual singularity trajectories, building a compact and permutation-free description of the collective vortex evolution upon propagation.
%%%%%%%%%%%%%%%%%%%%%%%%%%%%%%% Figure 1 
\begin{figure}[!htbp]
\includegraphics[width=1\linewidth]{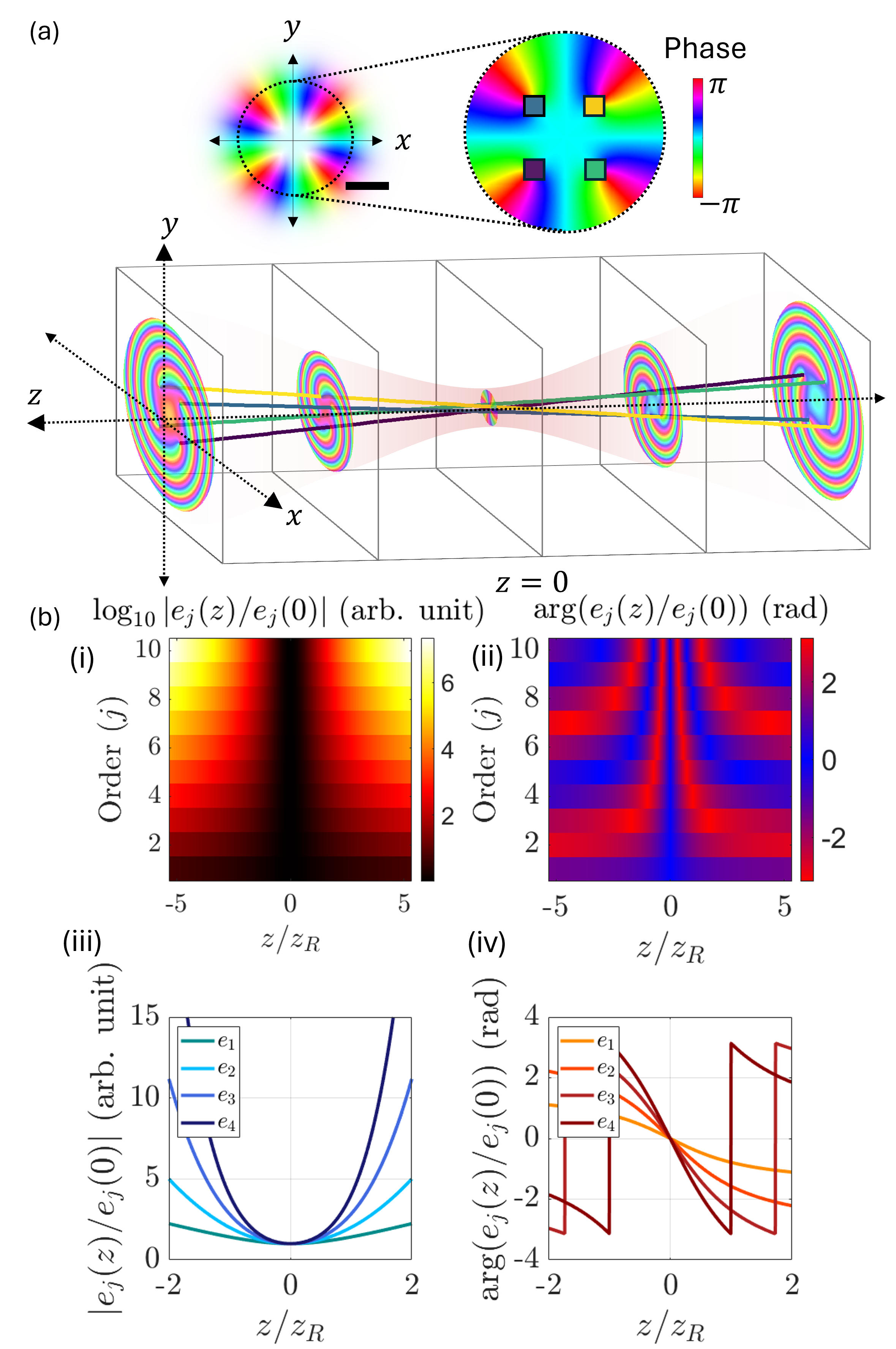}% Here is how to import EPS art
\caption{\label{fig:schematic}\textbf{Longitudinal propagation of an optical vortex constellation described by the complex elementary symmetric polynomials (ESPs) of the singularity coordinates.} 
(a) Schematically illustrated longitudinal propagation of a representative optical vortex constellation consisting of four unit charge singularities at  $u_1=(400, 400)\, \mu\text{m}$, $u_2=(400 , -400)\, \mu\text{m} $, $u_3=(-400, 400)\, \mu\text{m} $, and $u_4=(-400, -400)\, \mu\text{m} $ marked with colored boxes embedded on a Gaussian background of beam waist $1$ mm.
Scalebar: $800~\mu$m.
The numerically simulated longitudinal propagation shows a rotation of singularities around the optical axis when the field passes through the waist.
(b) Evolution of the ESPs along longitudinal propagation.
Evolution of the logarithmic amplitude (i) and 
phase (ii) of $e_j(z)/e_j(0)$ for orders $j = 1$ to $10$ obtained by Eq.~\eqref{eq:esp_prop}. 
Amplitude (iii) and phase (iv) evolution for the first four ESP orders within the proximity of the waist plane.}
\end{figure}
%%%%%%%%%%%%%%%%%%%%%%%%%%%%%%%%%%%%%%%%%%%%%%%%%%%%%
%
In this work, we introduce a simple law for the longitudinal propagation of an arbitrary vortex constellation  based on the joint spatial positions of all singularities, i.e., their distribution. 
We show that the entire longitudinal evolution of any optical vortex constellation formed by $n$ vortices can be described by a set of $n$ simple equations if the constellation is expressed in terms of complex elementary symmetric polynomials (ESPs) of the constituent singularity coordinates. 
We further experimentally verify this result for vortex constellations up to order four through tracing the evolution of the ESPs using off-axis digital holography techniques. 
The existence of such a simple deterministic propagation law for ESPs reveals a previously unrecognized mathematical symmetry in the longitudinal propagation, invariant to diverse vortex constellation geometries.
Importantly, beyond providing a compact representation of singularity propagation, we further demonstrate that the complex ESPs serve as a powerful metric for longitudinal metrology. 
We show that the higher the order of the ESPs, the more information they contain about the longitudinal position, i.e., the better they are for longitudinal sensing. 
Additionally, the amplitudes of the ESPs give the most information away from the waist region, whereas their phases exhibit maximal sensitivity in the vicinity of the focus.
These results establish ESP-based description of the longitudinal propagation of optical vortex constellations as a new route towards longitudinal position estimation based on singular optics.
%%%%%%%%%%%%%%%%%%%%%%%%%%%%%%%%%%%%%%%%%%%%%%%%%%%%%%%%%%%%%%%%%%%%%%%%%%%%%%%%%%%

%%%%%%%%%%%%%%%%%%%%%%%%%%%%%%%%%%%%%%%%%%%%%%%%%%%%%%%%%%%%%%%%%%%%%%%%%%%%%%%%%%%
\noindent \emph{Theoretical framework - }To start with, we consider a paraxial optical field propagating along the $z$ axis, containing, in the plane $z=0$, a vortex constellation of
$n=|\ell|$ unit-charge singularities embedded in a Gaussian background $E_0=E_0(z=0)$, i.e.,
\begin{equation}
E(u, z=0) = E_0 \prod_{k=1}^{n} ( u - u_k )
          = E_0 \sum_{j=0}^{n} (-1)^{j} e_j (0) \, u^{\,n-j},
\label{eq:fieldz0}
\end{equation}
where $u_k=x_k+iy_k$ is the transverse position of the singularity $k$, and $e_j$ denotes the $j$th elementary symmetric polynomial (ESP) of the complex vortex coordinates $u_k$, defined as
\begin{equation}
e_j = \sum_{1 \leq k_1 < k_2 < \cdots < k_j \leq n}
      u_{k_1} u_{k_2} \cdots u_{k_j},
\label{eq:espdef}
\end{equation}
with $e_0=1$. 
Fig. \ref{fig:schematic} (a) illustrates the propagation of such a constellation with four singularities.
Although expressions of this form have appeared implicitly in previous studies on propagation of vortex beams \cite{indebetouw1993optical}, the explicit identification of the coefficients $e_j$ as ESPs of the singularity coordinates, and their beneficial use as descriptors of the constellation geometries, has only been introduced recently \cite{barros2024observation}.
In the present work, the complex ESPs serve as the central parameters for a collective description of the longitudinal evolution of a vortex constellation, unlike in previous studies where the singularities were described individually.
Interestingly, ESPs, up to first few orders, can be interpreted as geometrical parameters of the constellation, i.e., $e_1$, $e_2$ and $e_3$ represent the barycenter, variance and skewness of the constellation respectively \cite{barros2024observation}.

%%%%%%%%%%%%%%%%%%%%%%%%%%%%%%%%%%%%%%%%%%%%% Figure 2
\begin{figure*}[t!]
\includegraphics[width=1\linewidth]{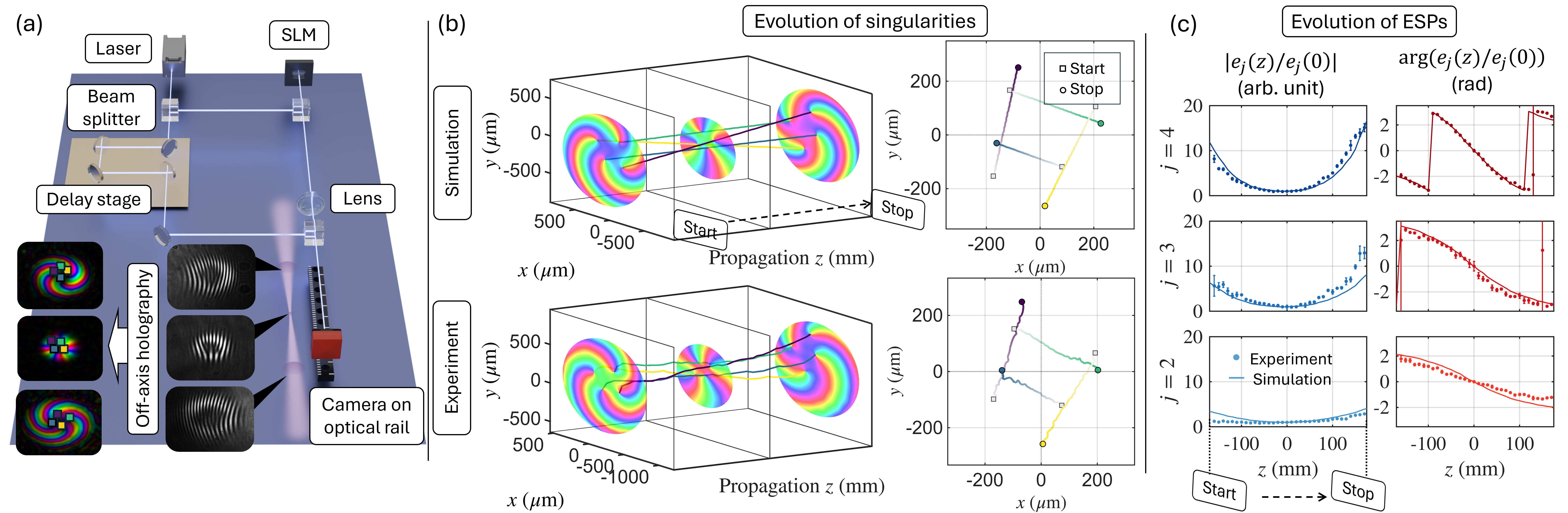}
\caption{\label{fig:expt}\textbf{Experimental observation of the propagation dynamics of vortex constellation described by the corresponding ESPs.} 
(a) Sketch of the experimental setup. 
A laser is split into two paths, with one being directed on a spatial light modulator (SLM) to generate the field containing optical vortex constellations using holographic methods. 
The second beam is adjusted in length via a delay stage, brought to interference with the modulated beam along the longitudinal propagation.
At each step, the resulting off-axis interferograms are recorded using a camera (see grey scale insets) and used to reconstruct the complex field and its singularity positions (see phase patterns in hue-color with singularities marked by squares.)
(b) The simulated and experimentally captured longitudinal propagation of the constellation depicted in (a) show a similar rotation %(apart from an angular offset caused by the mounting of the camera) 
when the field passes through waist (left column).
The rotation is more apparent in the stereographic projection of the trajectories of the singularities on the $x$-$y$-plane (right column).
(c) Evolution of ESPs along the propagation of the light field. 
The amplitude (left column) and phase (right column) evolutions of $e_j(z)/e_j(0)$ with $j=2,3,4$ obtained from experiment (circles) and simulation (lines) showing good agreement with each other. 
The error bars correspond to five times the standard deviation calculated from 20 measurements.}
\end{figure*}
%%%%%%%%%%%%%%%%%%%%%%%%%%%%%%%%%%%%%%%%%%%%%%%%%%%%%%%%%%%%

After free-space propagation to a distance $z$, the field becomes
\begin{equation}
E(u, z) \simeq  E_0(z) \zeta(z)^{n+1}\sum_{j=0}^{n} [-\zeta(z)]^{-j} e_{j}(0)\ u^{n-j},
\label{eq:fieldz}
\end{equation}
where $\zeta(z)=z_R/(z_R-iz)$, and $z_R$ is the Rayleigh range of the Gaussian background.
More details in the calculation can be found in \cite{SM}.

The common $z$-dependent prefactor outside the sum in Eq. \eqref{eq:fieldz} corresponds to the propagation factor of a Gaussian spatial mode carrying an on-axis vortex of charge $n$ ($\pm\ell$) and does not affect the vortex constellation.
Importantly, inside the summation the ESPs also evolves with $z$, obeying the simple propagation law 
\begin{equation}
e_j(z) = e_j(0)\zeta(z)^{-j}=e_j(0)\left( \frac{z_R}{z_R - i z} \right)^{-j}.
\label{eq:esp_prop}
\end{equation}

Eq.~\eqref{eq:esp_prop} establishes a universal paraxial propagation law independent of the number, arrangement, or shape of the vortex constellation, serving as the foundation of this work and specifically, for subsequent demonstration of longitudinal sensing.
Fig.~\ref{fig:schematic} (b) illustrates the evolution of ESPs $e_j(z)/e_j(0)$ of different order $j$ with changing $z$.
Note that such evolution of ESPs does not cover background fields with nonzero radial structure, e.g., higher order radial modes.
In such scenarios, vortex-antivortex pairs are generated during longitudinal propagation, which can result in vortex links and knots \cite{leach2004knotted,dennis2010isolated}. 
Albeit interesting, it is outside the scope of this work if and how EPSs can be leveraged to find a simplified depiction of the complex dynamics for such scenarios. 
%%%%%%%%%%%%%%%%%%%%%%%%%%%%%%%%%%%%%%%%%%%%%%%%%%%%%%%%%%%%%%%%%%%%%%

%%%%%%%%%%%%%%%%%%%%%%%%%%%%%%%%%%%%%%%%%%%%%%%%%%%%%%%%%%%%%%%%%%%%%%
\noindent\emph{Experimental verification - }To verify the introduced ESP formalism for the description of the propagation dynamics of vortex constellations,  we experimentally track the evolution of the corresponding ESPs extracted from the singularity coordinates (see Fig.~\ref{fig:expt}).
A continuous-wave laser ($\lambda=810$~nm) is spatially filtered and collimated to approximate a fundamental Gaussian mode.
The beam is divided into two arms of a Mach--Zehnder interferometer one of which has a spatial light modulator (SLM) imprinting vortex constellations with desired positions of unit-charge phase singularities (see Fig.~\ref{fig:expt} (a)) using holographic methods \cite{forbes2016creation}.
The modulated beam is subsequently focused by a lens of focal length $f=500$~mm, thereby realizing free-space paraxial propagation within an experimentally accessible distance.
The second arm serves as a reference for subsequent off-axis digital holography \cite{verrier2011off}, which enables the retrieval of the complex field and, therefore, the precise determination of the vortex positions.
To save the holograms from contamination of unwanted interference arising from back reflections, the laser is operated below the lasing threshold.
The interferograms are recorded by a camera mounted on an optical rail, allowing measurements over the range $z\in[-160,170]$~mm, where $z=0$ denotes the focal plane of the lens mimicking the waist of free space propagation.

We start with studying a representative vortex constellation with four unit charge singularities located at $u_1=(400, 400)\, \mu\text{m}$, $u_2=(200 , -200)\, \mu\text{m} $, $u_3=(-400, 400)\, \mu\text{m} $, and $u_4=(-400, -400)\, \mu\text{m} $ in the complex plane $(x, y)$ with a Gaussian background centered at $(0,0) $ having a beam waist of $700$~$\mu$m. 
The camera is moved along the optical rail in steps of $10$~mm while capturing the corresponding interferogram at each position. 
The singularity positions at different longitudinal planes are located by the subsequent off-axis digital holography. 
From these spatial positions, the ESPs are calculated using Eq.~\eqref{eq:espdef}. 
For achieving experimental stability in determining singularity positions and therefore the ESPs in the full range of $z$, the singularity positions, for every reconstructed field, are recalculated from the barycenter of the constellation, i.e., the first ESP $e_1$. 
This revaluation automatically sets $e_1= 0$, while all other higher order ESPs remain intact.
Fig.~\ref{fig:expt} (b) shows a comparison between experimentally recorded and corresponding numerically simulated longitudinal propagation of the constellation.
The corresponding 2D stereographic projection on $x-y$ plane reveals the movement of the singularities.
The slight rotational offset in the experimental projection compared to the simulation is an experimental artifact due to the imprecise angle adjustment when mounting the camera. 
We restrained from adjusting this offset in postprocessing as digital rotations introduce additional errors and the evolution of the ESPs ($e_j(z)/e_j(0)$) is insensitive to such rotational offsets.
Fig.~\ref{fig:expt} (c) presents the experimentally observed longitudinal evolution of the complex ESPs $e_2(z)/e_2(0), e_3(z)/e_3(0), e_4(z)/e_4(0)$.
The minor mismatch in experimental and simulated ESP evolution stems from small imprecision in estimating the waist position of the beam.
However, it is evident that ESPs provide a useful handle to describe an optical vortex constellation and their evolution can successfully express the longitudinal propagation of a vortex constellation.
Note that we performed experiments with different structure of vortex constellation containing different number of singularities and such evolution of ESPs is invariant (see \cite{SM}).
%%%%%%%%%%%%%%%%%%%%%%%%%%%%%%%%%%%%%%%%%%%%%%%%%%%%%%%%%%%%%%%%%%%%%%%%

%%%%%%%%%%%%%%%%%%%%%%%%%%%%%%%%%%%%%%%%%%%%%%%%%%%%%%%%%%%%%%%%%%%%%%%%

%%%%%%%%%%%%%%%%%%%%%%%%%%%%%%%%%%%%%%%%%%%%%%%%  Figure 3
\begin{figure}[h]
\includegraphics[width=1\linewidth]{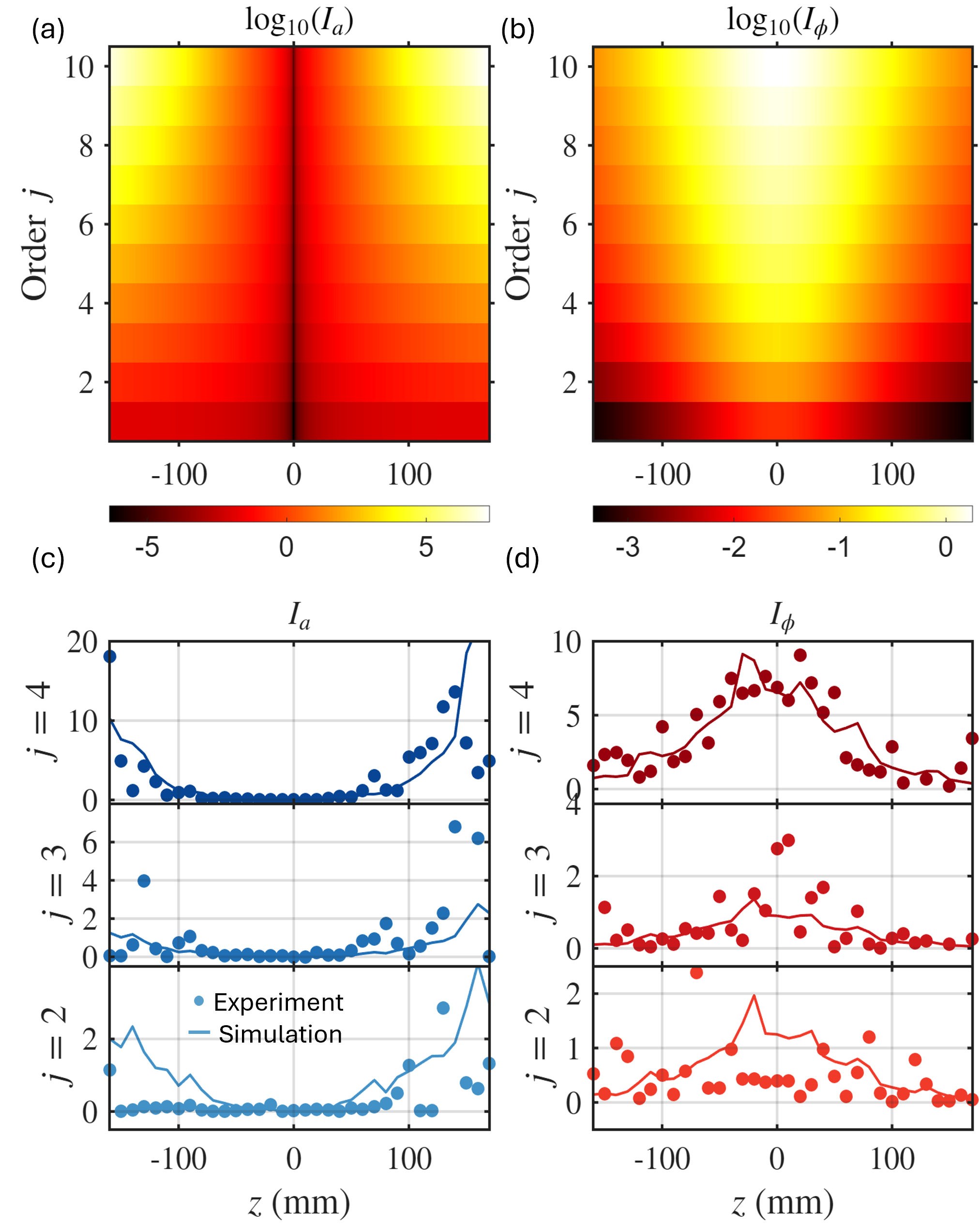}
\caption{\label{fig:fisher}\textbf{Capability of ESPs $e_j(z)/e_j(0)$ in longitudinal length sensing demonstrated by corresponding Fisher information.} 
(a), (b) Theoretical variation of Fisher information $\mathcal{I}_a$ and $\mathcal{I}_\phi$ respectively with varying $z$ and $j$ considering Rayleigh range to be $75$ mm. $\log_{10}$ scale has been used to squeeze the scale for more apparent visual representation.
(c), (d) Experimentally obtained variation of $\mathcal{I}_a$ and $\mathcal{I}_\phi$ respectively (circles), with changing longitudinal distance $z$ along with corresponding simulation prediction (lines) for different ESP order $j= 2\ , 3\ , 4$.}
\end{figure}
%%%%%%%%%%%%%%%%%%%%%%%%%%%%%%%%%%%%%%%%%%%%%%%%%%%%%%%%%%%%%%

\noindent \emph{ESP for longitudinal sensing - }The ESP-based description of the propagation of vortex beams opens up a new route for sensing longitudinal displacement.
Interestingly, the amplitude of ESPs  $|e_j(z)/e_j(0)|$ shows a stronger variation along $z$ away from the waist region, i.e., $z>z_R$, while it tends to saturate ($|e_j(z)/e_j(0)|\rightarrow 1,\forall j$) at the beam waist $z=0$.
In contrast, the phase of the ESP $\arg(e_j(z)/e_j(0))$ shows a larger variation with respect to the distance $z$ indicating a better sensitivity when sensing around the focal region is desired. 
In both cases, this sensitivity increases the higher the order of the ESP $j$, showing that more complex constellations allow for better longitudinal sensing (see Fig.~\ref{fig:expt} (c)). 
We further quantify the longitudinal sensing ability when using ESPs by calculating the resulting Fisher information $\mathcal{I}$ when measuring the longitudinal evolution of the amplitude ($\mathcal{I}_a$) and phase ($\mathcal{I}_\phi$) part of $j$-th ESP.
The amount of Fisher information obtainable in any measured parameter $S(z)$ to estimate $z$ is $\mathcal{I}=E\left[\frac{1}{\sigma^2}\left(\frac{dS}{dz}\right)^2\right]
\label{eq:fisher_basic}$ where $\sigma^2$ is the variance of the additive measurement noise on
$S(z)$, assumed normally distributed~\cite{kay1993statistical}.
Using Eqs.~\eqref{eq:esp_prop}, we obtain the Fisher information in our case.
\begin{equation}
\mathcal{I}_a^{(j)}=\frac{j^2z^2}{\sigma_{j,a}^2z_R^4}\left(1+\frac{z^2}{z_R^2}\right)^{j-2};\ \mathcal{I}_\phi^{(j)}=\frac{j^2z_R^2}{\sigma_{j,\phi}^2(z^2+z_R^2)^4}
\label{eq:fisher}
\end{equation}
where $\sigma_{j,a/\phi}^2$ is the variance in the experimental estimation of the amplitude/phase part of ESPs.  
Eq.~\eqref{eq:fisher} confirms that $\mathcal{I}_a$ and $\mathcal{I}_\phi$ become significant in complementary regions of $z$.
$\mathcal{I}_a^{(j)}$ vanishes at the focal plane $z=0$, i.e., no information about the longitudinal position can be deduced, and gradually increases when moving away from the focus.
In contrast, the Fisher information given by the phase evolution $\mathcal{I}_\phi^{(j)}$ has its maximum at $z=0$, and decreases when moving away from this plane. 
Therefore, the complex ESPs $e_j(z)/e_j(0)$ is capable of sensing longitudinal distance $z$ in the full domain of longitudinal distance $z$.
To demonstrate such complementary behaviour and to emphasize the enhancement of Fisher information with increasing $j$, we consider a common $\sigma_{j,a/\phi}= 0.1$ for all $j$ and $z_R=75\ \text{mm}$ and plot $\mathcal{I}_a$ and $\mathcal{I}_\phi$ with varying $z$ and $j$ are shown in Fig. \ref{fig:fisher} (a) and (b) respectively. 

We compare these results to our experiment using the longitudinal propagation of the vortex constellation presented in Fig. \ref{fig:expt}, from which we calculate both Fisher information $\mathcal{I}_a$ and $ \mathcal{I}_\phi$.
Note that, unlike previous theoretical considerations, the noise $\sigma_j$ for different $j$ actually varies over $z$. 
This is because the background field changes rapidly near the focus, reshaping the off-axis holograms and thus varying the uncertainty in locating the singularities.
We average $\sigma_j$ over $z$ and take the mean value as $\sigma_j$ to estimate the Fisher information.
For experimental $\mathcal{I}_a$ and $\mathcal{I}_{\phi}$, we find the mean values $\sigma_{j,a} = \{0.0183, 0.0618, 0.052\}$ and $\sigma_{j,\phi} = \{0.0141, 0.0286, 0.0162\}$, for $j=\{1,2,3\}$ respectively.
The Fisher information is also calculated from simulations with the above mentioned experimental parameters.
Fig. \ref{fig:fisher} (c), (d) depicts the variation of $\mathcal{I}_a$ and $\mathcal{I}_\phi$  with changing $j$ and $z$ estimated from experiment and simulation, which shows an agreement between them.
For lower-order ESPs ($j=2,3$), the experimentally estimated Fisher informations, specifically $\mathcal{I}_a$, deviate more significantly from simulations because their inherently smaller magnitudes are more easily overwhelmed by experimental fluctuations.
In contrast higher order ESPs ($j=4$) agree much better with our simulations which nicely demonstrates the increased information gain when using constellations of a larger number of vortices, i.e., their superiority in terms of longitudinal sensing.
These results confirm that ESPs of an optical vortex constellation provide a platform to use the optical singularities in longitudinal length sensing. 
%%%%%%%%%%%%%%%%%%%%%%%%%%%%%%%%%%%%%%%%%%%%%%%%%%%%%%%%%%%%%%%%%%

%%%%%%%%%%%%%%%%%%%%%%%%%%%%%%%%%%%%%%%%%%%%%%%%%%%%%%%%%%%%%%%%%%

\noindent \emph{Conclusions - }
In summary, we have formulated and experimentally demonstrated a compact propagation law for arbitrary optical vortex constellations that eliminates the need to track individual singularities one by one.
As a practical benefit, since vortices are intensity nulls, the demonstrated law yields the phase profile and OAM spectrum from intensity images alone, without direct phase measurement.
Due to the universal nature of our results, our theoretical formalism of the propagation of vortex constellation might also be applicable to Bose–Einstein condensates \cite{butts1999predicted}, superfluids \cite{salomaa1987quantized} and topological field theories \cite{witten1988topological} to describe evolution of singularities building further connections between these fields and optical vortex beams.

Beyond the fundamental importance, we have shown that this ESP-based description yields a powerful tool for precision longitudinal sensing. 
Through Fisher information analysis, we demonstrated that complex ESPs inherently encode longitudinal distance information, with their amplitude and phase components exhibiting maximal sensitivity in complementary domains of $z$. 
Furthermore, higher-order ESPs possess greater sensitivity to propagation, making complex constellations inherently superior for longitudinal length estimation. 
This establishes a new route toward longitudinal sensing through singularimetry.

We have demonstrated this ESP-based description in the most common and fundamental scenario of scalar spatial vortex constellations propagating in free space.
The formalism immediately extends to the constellation of polarization singularities, e.g., the C-points of a vector beam which are vortices of its circular polarization component of the field \cite{nye1983lines}.
Further, the demonstrated foundation paves the way for incorporating more complex scenarios, such as spatio-temporal optical vortices \cite{bliokh2012spatiotemporal} and vortex propagation through nonlinear media \cite{luther1994nonlinear,rozas2000observed}.
Additionally, the formalism can be extended for vortex constellation embedded on a non-Gaussian background, i.e., with nonzero radial order to incorporate vortex links and knots, as mentioned earlier.
We believe that our work finds fundamental interest from different branches of physics working with singularity and secures application in optical metrology and sensing.

\begin{acknowledgments}
Authors acknowledge the helpful discussions with Matias Eriksson, Jaime Moreno Zuleta, and Sophia Strnat in different aspects in theory, simulation and experiment. 
NM and MO acknowledge the financial support from the Photonics Research and Innovation Flagship (PREIN - Decision 346511). 
RF acknowledges the support of the Research Council of Finland through the Academy Research Fellowship (decision 332399) and through the project BIQOS (decision 358134), and  European Research Council (ERC) Starting grant TWISTION (101042368). RB acknowledges the financial support of FAPESP (grant 2024/08450-0).
\end{acknowledgments}

\bibliography{apssamp}

%%%%%%%%%%%%%%%%%%%%%%%%%%%%%%% Supplementary %%%%%%%%%%%%%%%%%%%%%%%%%%

% ==============================================================================
% =========================== SUPPLEMENTAL MATERIAL ============================
% ==============================================================================

\clearpage
\onecolumngrid
\begin{center}
\textbf{\large Supplementary information: A Collective Propagation Law of Optical Vortex Constellations and Longitudinal Sensing}
\end{center}

% ---------- Prefix an "S" to all equations, figures, tables and reset the counters ----------
\setcounter{equation}{0}
\setcounter{figure}{0}
\setcounter{table}{0}
\setcounter{page}{1}
\makeatletter
\renewcommand{\theequation}{S\arabic{equation}}
\renewcommand{\thefigure}{S\arabic{figure}}
\renewcommand{\bibnumfmt}[1]{[S#1]}
%\renewcommand{\citenumfont}[1]{S#1}
% -------------------------------------------------------------------------------------------

\section{Theoretical formalism}
This section presents the theoretical formulation for the longitudinal propagation of an optical vortex constellation in terms of the evolution of elementary symmetric polynomials (ESPs) of the singularity coordinates.
We consider a paraxial optical field containing a constellation of $n$ unit-charge singularities embedded in a Gaussian background $E_0$ centered at $(0, 0)$ on the transverse $(x,y)$ plane having a beam waist $w_0$.
At the waist plane $z=0$, the complex field can be written as

\begin{equation}
E(u, z=0) = E_0 \prod_{k=1}^{n} ( u - u_k )
          = E_0 \sum_{j=0}^{n} (-1)^{j} e_j (z=0) \, u^{\,n-j},
\label{supp_eq:fieldz0}
\end{equation}
where $\{u_1,\ldots,u_n\}$ denote the transverse positions of the unit-charge vortices in the complex plane $u=x+iy$.
The coefficients $e_j$ fully encode the geometry of the vortex constellation.

Equation~\eqref{supp_eq:fieldz0} further shows that a constellation of $n$ unit-charge vortices can be interpreted as a coherent superposition of on-axis vortices with topological charges ranging from $0$ to $n$, weighted by the coefficients $e_j$, the ESP of order $j$ of the constituent vortex coordinates.
This approach has been previously mentioned in \cite{indebetouw1993optical}; however, the identification of $e_j$ as the ESP of the vortex coordinates and using them to describe an optical vortex constellation has been done only very recently \cite{barros2024observation}.
Mathematically, $e_j$ can be calculated from the singularity coordinates as 

\begin{equation}
e_j = \sum_{1 \leq k_1 < k_2 < \cdots < k_j \leq n}
      u_{k_1} u_{k_2} \cdots u_{k_j},
\end{equation}
with $e_0=1$.

To evaluate the propagated field \(E(u,z)\), it is convenient to work in polar coordinates
\(u=re^{i\theta}\).
Up to an overall normalization constant \(N\), the field at the waist plane \(z=0\) becomes

\begin{equation}
E(r,\theta,0)=
N \exp\!\left(-\frac{r^2}{w_0^2}\right)
\sum_{j=0}^{n}(-1)^j e_j(0)\,(re^{i\theta})^{\,n-j},
\label{eq:field_polar_0}
\end{equation}
where $e_j(0)\equiv e_j(z=0)$

Free-space paraxial propagation is described in the spatial-frequency domain by the transfer
function
$H(\rho,z)\sim\exp(i\pi\lambda z\rho^2)$,
with $\rho$ being the transverse spatial frequency and $\lambda$ being the wavelength.
The angular spectrum at the waist $\mathcal{E}(\rho,\phi,0)=\mathcal{F}\{E(r,\theta,0)\}$,  using the Fourier transform identity for a helical Gaussian mode $\mathcal{F}\!\left\{(re^{i\theta})^{m}e^{-r^2/w_0^2}\right\}=\pi w_0^2
\big(i\pi w_0^2\rho e^{i\phi}\big)^{m}
e^{-\pi^2 w_0^2\rho^2}$, can be obtained as \cite{goodman1969introduction},
\begin{equation}
\mathcal{E}(\rho,\phi,0)=
N \pi w_0^2 e^{-\pi^2 w_0^2\rho^2}
\sum_{j=0}^{n}(-1)^j e_j (0)\ 
\big(i\pi w_0^2\rho e^{i\phi}\big)^{\,n-j}.
\end{equation}
After multiplication by the transfer function \(H(\rho,z)\), the propagated angular spectrum becomes
\begin{equation}
\mathcal{E}(\rho,\phi,z)=
N \pi w_0^2
\sum_{j=0}^{n}(-1)^j e_j (0)\ 
\big(i\pi w_0^2\rho e^{i\phi}\big)^{\,n-j}
\exp\!\left[(-\pi^2 w_0^2+i\pi\lambda z)\rho^2\right].
\end{equation}
The spatial field $E(r,\theta,z)$ can be recovered by an inverse Fourier transformation.
Introducing the complex beam parameter \(\bar{w}^2(z)=w_0^2\!\left(1-i z/z_R\right)\), with Rayleigh range \(z_R=\pi w_0^2/\lambda\), the spatial field is obtained as
\begin{equation}
E(r,\theta,z)=
\frac{N w_0^2}{\bar{w}^2(z)}
\exp\!\left(-\frac{r^2}{\bar{w}^2(z)}\right)
\sum_{j=0}^{n}(-1)^j e_j(0)\ 
\left(\frac{w_0^2}{\bar{w}^2(z)}\,re^{i\theta}\right)^{\,n-j}.
\label{eq:field_propagated}
\end{equation}
Returning to the complex transverse coordinate \(u=re^{i\theta}\) and defining the dimensionless
propagation factor
\begin{equation}
\frac{1}{\zeta(z)}=1-\frac{i z}{z_R},
\end{equation}
the propagated field can be compactly written as
\begin{equation}
E(u,z)=
\frac{N}{S(z)}
\exp\!\left(-\frac{|u|^2 \zeta(z)}{w_0^2}\right)
\sum_{j=0}^{n}(-1)^j e_j\,
\frac{u^{\,n-j}}{\zeta(z)^{\,j-n}} .
\label{eq:final_prop}
\end{equation}
Identifying $E_N \exp(-\frac{|u|^2 \zeta(z)}{w_0^2}) = E_0(z)$ and by factoring out the common power \(\zeta(z)^{n}\), Eq.~\eqref{eq:final_prop} can be cast in the form
\begin{equation}
E(u,z)=
E_0(z) \zeta(z)^{\,n+1}
\sum_{j=0}^{n}(-1)^j
\big[\,e_j(0)\, \zeta(z)^{-j}\big]\,
u^{\,n-j}.
\end{equation}
Now if we look inside the summation, we find that the ESP $e_j$ which describes the constellation structure has also evolved at $z$
\begin{equation}
e_j(z)=e_j(0)\,\zeta(z)^{-j}
      =e_j(0)\left(\frac{z_R}{z_R-i z}\right)^{-j},
\end{equation}
giving rise to the propagation law mentioned in Eq. (4) of the main text. 

The derivation above explicitly describes a constellation of positive unit-charge singularities ($+1$). 
For a constellation composed of negative unit-charge singularities ($-1$), the local azimuthal phase gradient is reversed. 
Mathematically, this is accounted for by taking the complex conjugate of the transverse coordinates, such that $u \rightarrow u^* = r e^{-i\theta}$. 
Consequently, the constituting ESPs evaluated at the waist become complex conjugated, $e_j \rightarrow e_j^*$. Propagating this conjugated field through the Fourier domain modifies the standard helical Fourier transform identity, yielding $\mathcal{F}\{(re^{-i\theta})^{m}e^{-r^2/w_0^2}\} \propto (-i\rho e^{-i\phi})^{m}$. 
Multiplication with paraxial transfer function and subsequent inverse Fourier transform yields a complementary propagation law for negative vortex constellations.

\begin{equation}
e_j(z)=e_j(0)\left(\frac{z_R}{z_R+i z}\right)^{-j}.
\end{equation}

Physically, this change from $(z_R - iz)$ to $(z_R + iz)$ for vortex constellation of $-1$ charged singularities manifests in the reverse evolution of only $\arg(e_j(z)/e_j(0))$ to that for $+1$ charged singularities.

\section{Propagation of different vortex constellation}

\begin{figure*}[t!]
\includegraphics[width=\linewidth]{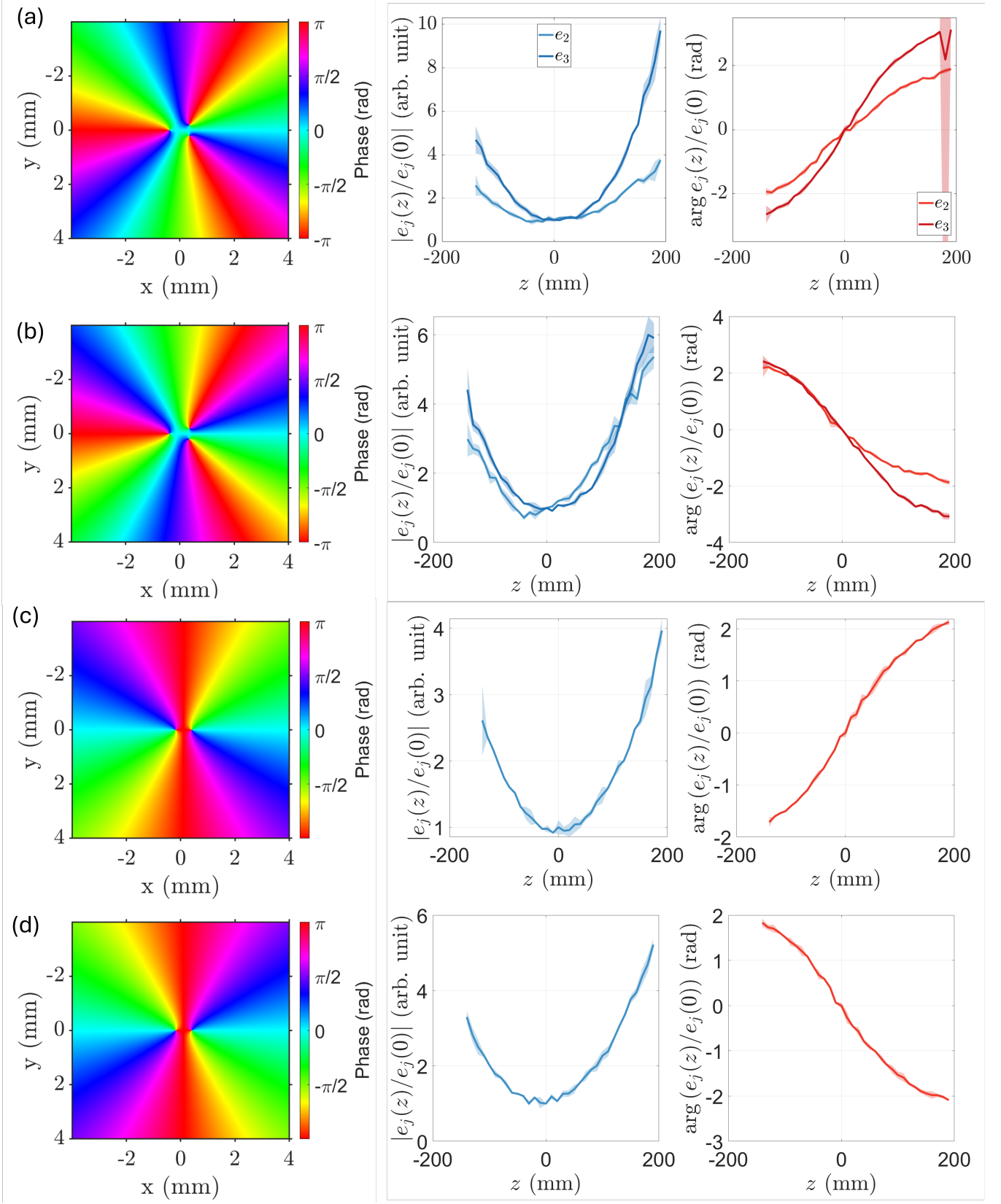}
\caption{\textbf{Longitudinal evolution of arbitrarily different optical vortex constellations captured by the evolution of ESPs.} Different constellations are generated on same Gaussian background centered at $(0,0)$ on $(x,y)$ plane with three (a, b) and two (c, d) unit charge singularities in the same experimental arrangement mentioned in the main text. (a), (b) For constellations with three unit charge singularities, we only have nonzero $e_2$ and $e_3$ when the singularity coordinates are recalculated from their barycenter. When the constellation contains three $-1$ charged singularities (a), the evolution of $\arg(e_j(z)/e_j(0))$ is flipped with respect that of constellation of three three $+1$ charged singularities (b). (c), (d) For a constellation two unit charged singularities, we find the expected behaviour in the experimental evolution of ESPs, however, now only $e_2$ exists. The shaded errorbars in all the experimental plots correspond to five times the standard deviation calculated from 20 measurements.}
\label{fig:supp_ESPs}
\end{figure*}

We performed the experiment for observing propagation of different vortex constellation through the evolution of ESPs containing different number of unit charged singularities and also with flipping the sign of unit charge ($+1\rightarrow -1$). Note that for such sign flip, the evolution of $|e_j(z)/e_j(0)|$ remains same, though the evolution of $\arg(e_j(z)/e_j(0))$ flips, as also mentioned in the previous section (see Fig.~\ref{fig:supp_ESPs}).

\end{document}